\documentclass[aps,twocolumn]{revtex4}
\usepackage{epsfig}
\usepackage{graphicx}
\usepackage{amsmath}
\usepackage{booktabs}
\usepackage{setspace}
\begin{document}
\begin{spacing}{1.0}
\title{Decoding the double heavy tetraquark state $T^+_{cc}$}

\author{Cheng-Rong Deng$^{a,b}$
and Shi-Lin Zhu$^{b}{\footnote{zhusl@pku.edu.cn}}$}
\affiliation{$^a$School of Physical Science and Technology,
Southwest University, Chongqing 400715, China}
\affiliation{$^b$School of Physics and Center of High Energy
Physics, Peking University, Beijing 100871,China}

\begin{abstract}

We analyse the deuteron-like $T^+_{cc}$ state observed by the LHCb
Collaboration from the perspective of the quark clustering. The
$T^+_{cc}$ state implies the possible existence of other compact
doubly heavy tetraquark states.

\end{abstract}
\maketitle

Recently, the LHCb Collaboration reported the doubly charmed state
$T^+_{cc}$ with $IJ^P=01^+$ in the $D^0D^0\pi^+$ invariant mass
spectrum~\cite{tcc-exp}. Its extremely small binding energy and
width can match very well with the prediction of the $DD^\ast$
molecular state~\cite{tcc-lee,tcc-lmeng}. The deuteron-like molecule
configuration is manifest from its characteristic size. The
discovery of the $T^+_{cc}$ is a great breakthrough in hadron
physics after the state $\Xi^{++}_{cc}$, which provides an excellent
platform to investigate the low energy strong interactions.

Inspired by the $T^+_{cc}$, we employed the quark model to
systematically investigate the double heavy tetraquark states with
the molecule configuration~\cite{deng2022}. We extracted the binding
energy of the $T^+_{cc}$ to be 0.34 MeV, which agrees with LHCb's
measurement very well. The $T^+_{cc}$ is a loosely bound
deuteron-like state with a huge size around 4.32 fm. The long-range
$\pi$ and intermediate-range $\sigma$ exchange interactions and
coupled channel effect play a pivotal role. In the molecule
configuration $[b\bar{q}]_{\mathbf{1}}[b\bar{q}]_{\mathbf{1}}$, the
$T^-_{bb}$ state with $01^+$ is sensitive to different dynamical
effects and may have three different physical pictures: the compact
state, deuteron-like state, or hydrogen moleculelike state. Their
binding energies are less than 50 MeV and qualitatively consistent
with the latest lattice QCD predictions~\cite{bicudo2021}.

Prior to the $T^+_{cc}$, many theoretical explorations have focused
on the stability of the doubly heavy tetraquark states since the
pioneering work~\cite{ader1982}. The $T^-_{bb}$ state with $01^+$
was widely accepted as the most promising stable doubly heavy
tetraquark state with a binding energy around $-200$ MeV relative to
the $\bar{B}\bar{B}^*$ threshold~\cite{deng2020}. However, the
energy of the $T^+_{cc}$ with $01^+$ hovers over the $DD^*$
threshold in the range of 300 MeV. Its stability is highly model
dependent.

In the nonrelativistic quark models, the diquark configuration
$[QQ][\bar{q}\bar{q}]$ ($Q=c$ and $b$, $q=u$ and $d$) is apt to
yield a deep compact bound state while the molecule configuration
$[Q\bar{q}][Q\bar{q}]$ is prone to produce a shallow bound state.
One can consult the latest review on the doubly heavy tetraquark
states~\cite{chen2022,meng2022}. In this work, we attempt to decode
the underlying mechanism of these two obviously different physical
pictures from the perspective of the quark clustering.

Gell-Mann first proposed the possibility of the diquark in his
pioneering work in 1964~\cite{gell-mann1964}. It seems that there
exists some phenomenological evidence of the relevance of the
diquarks in hadron
physics~\cite{jaffe2005,anselmino1993,barabanov2021}. We want to
emphasize that the diquark is not a point-like fundamental object as
the quark. Throughout this work, we use the diquark to denote the
possible quark clustering and correlations. In fact, the diquark is
a spatially extended object with various color-flavor-spin-space
configurations. The substructure of the diquarks may affect the
structure of the multiquark states.

We define the color quantum number $c=0$ and $c=1$ for the diquark
in the color $\bar{\mathbf{3}}$ and $\mathbf{6}$ representation
respectively. For the heavy diquark $[QQ]$, its spin $s$, isospin
$i$, orbit angular excitation $l$, and color $c$ should satisfy the
constraint $s+i+l+c=odd$ due to the Pauli principle. The $S$-wave
diquark $[QQ]_{\bar{\mathbf{3}}}$ must be spin triplet. Its Coulomb
interaction is strongly attractive because the large mass of the
heavy quarks decreases the kinetic energy and allows them to
approach each other. The heavier the heavy quark, the stronger the
Coulomb interaction. Its color-magnetic interaction is weakly
repulsive because it is suppressed by the heavy quark mass. Thus,
the $S$-wave diquark $[QQ]_{\bar{\mathbf{3}}}$ is favored.

The $S$-wave diquark $[QQ]_{\mathbf{6}}$ must be spin singlet. Both
the color-magnetic interaction and Coulomb interaction are repulsive
so that this type of the heavy diquark is disfavored. However, the
diquark-antidiquark configuration
$\left[[QQ]_{\mathbf{6}}[\bar{Q}\bar{Q}]_{\bar{\mathbf{6}}}\right]_{\mathbf{1}}$
is predominant in the fully-heavy teraquark ground states due to the
strong Coulomb attraction and confinement potential between the two
color sextet subclusters ~\cite{deng2021}. For the excited diquarks
$[QQ]_{\bar{\mathbf{3}}}$ and $[QQ]_{\mathbf{6}}$, their
color-magnetic interaction is weak. The Coulomb interaction
decreases because they are spatially more extended. The orbital
excitation tends to increase the kinetic energy. Thus, the excited
diquarks are heavier than the $S$-wave diquark.

For the light diquark $[qq]$, the situation is opposite to that of
the heavy diquark $[QQ]$ because of the obvious mass difference. The
color-magnetic interaction tends to prevail over the Coulomb
interaction in the diquark $[qq]$. Its spin $s$, isospin $i$, orbit
angular excitation $l$, and color $c$ obey the constraint
$s+i+l+c=even$. The $S$-wave diquark $[qq]$ has four possible
spin-isospin-color combinations. The spin singlet, isospin singlet
and color triplet diquark is often called the good diquark, which is
simultaneously favored by the Coulomb interaction, color-magnetic
interaction as well as the one-pion-exchange interaction. Other
combinations are sometimes called bad diquarks. A good diquark
$[qq]_{\bar{\mathbf{3}}}$ and a good anti-diquark
$[\bar{q}\bar{q}]_{\mathbf{3}}$ generally do not form a stable
tetraquark state because of the low mass threshold of two light
pseudoscalar mesons.

A good diquark $[QQ]_{\bar{\mathbf{3}}}$ and a good antidiquark
$[\bar{q}\bar{q}]_{\mathbf{3}}$ are an optimal combination to
produce a possible stable tetraquark state with $01^+$. The Coulomb
interaction in the diquark $[QQ]_{\bar{\mathbf{3}}}$ alone can
ensure that the doubly heavy tetraquark lies below the threshold of
two $Q\bar{q}$ mesons if the mass ratio of ${M_Q}$ and ${m_q}$
exceeds a critical value. In the limit of the very large ${M_Q}$,
the diquark $[QQ]_{\bar{\mathbf{3}}}$ shrinks into a tiny and
compact core because of the strong Coulomb interaction while the
light quarks move around the $QQ$-core~\cite{deng2022}. The doubly
heavy tetraquark states look like a helium-like QCD-atom. Moreover,
the color-magnetic interaction and pseudoscalar meson exchange force
are also very strong if one takes chiral symmetry into account in
the good antiquark $[\bar{q}\bar{q}]_{\mathbf{3}}$. To some extent,
the good antidiquark $[\bar{q}\bar{q}]_{\mathbf{3}}$ plays the
similar role of the electron pair in the hydrogen molecule, where
the two electrons are in spin singlet and form a $\sigma$-bond.
Furthermore, the above two binding mechanisms are independent and do
not occur in the threshold of two $Q\bar{q}$ mesons, which is
beneficial to produce the compact tetraquark states with $01^+$.

The doubly heavy tetraquark states have two completely equivalent color
configurations~\cite{vijande2009}:
\{$\mathbf{1}_{12}\otimes\mathbf{1}_{34}$,
$\mathbf{8}_{12}\otimes\mathbf{8}_{34}$\} for
$[Q_1\bar{q}_2][Q_3\bar{q}_4]$ and
\{$\bar{\mathbf{3}}_{13}\otimes\mathbf{3}_{24}$,
$\mathbf{6}_{13}\otimes\bar{\mathbf{6}}_{24}$\} for $[Q_1
Q_3][\bar{q}_2\bar{q}_4]$, where the subscript represents the quark index.
For the tetraquark states with $01^+$, their spin configurations are also
equivalent~\cite{vijande2009}:
\{$\mathbf{1}_{12}\oplus\mathbf{1}_{34}$,
$\mathbf{1}_{12}\oplus\mathbf{0}_{34}$,
$\mathbf{0}_{12}\oplus\mathbf{1}_{34}$\} and
\{$\mathbf{1}_{13}\oplus\mathbf{1}_{24}$,
$\mathbf{1}_{13}\oplus\mathbf{0}_{24}$,
$\mathbf{0}_{13}\oplus\mathbf{1}_{24}$\}. For the orbital space, one
can define the relative coordinates $\mathbf{r}_{ij}$ and
$\mathbf{r}_{ij,kl}$ as
\begin{eqnarray}
\begin{array}{cccccc}
\mathbf{r}_{ij}=\mathbf{r}_i-\mathbf{r}_j,~\mathbf{r}_{ij,kl}=\frac{m_i\mathbf{r}_i+m_j\mathbf{r}_j}{m_i+m_j}
-\frac{m_k\mathbf{r}_k+m_l\mathbf{r}_l}{m_k+m_l},\nonumber
\end{array}
\end{eqnarray}
where \{$\mathbf{r}_{12}$, $\mathbf{r}_{34}$, $\mathbf{r}_{12,34}$\}
and \{$\mathbf{r}_{13}$, $\mathbf{r}_{24}$, $\mathbf{r}_{13,24}$\}
are the Jacobi coordinates of the molecule configuration and the
diquark configuration, respectively. They can reciprocally transform
into each other through a unitary matrix depending on the quark
masses.

To obtain the diquark-antidiquark components in the molecular ground
state $[Q_1\bar{q}_2][Q_3\bar{q}_4]$, one can express the orbit wave
function $\phi(\mathbf{r}_{12},\mathbf{r}_{34},\mathbf{r}_{12,34})$
in terms of the explicit diquark configuration, namely
$\phi(\mathbf{r}_{12},\mathbf{r}_{34},\mathbf{r}_{12,34})
=\phi'(\mathbf{r}_{13},\mathbf{r}_{24},\mathbf{r}_{13,24})$. We
adopt the Gaussian function as the orbit trial wave function. The
coordinate-related parts can be expressed as
\begin{eqnarray}
e^{-\nu_i\mathbf{r}^2_{12}-\nu_j\mathbf{r}^2_{34}-\nu_k\mathbf{r}^2_{12,34}}
=e^{-a\mathbf{r}^2_{13}-b\mathbf{r}^2_{24}-c\mathbf{r}^2_{13,24}}\nonumber\\
\times e^{-d\mathbf{r}_{13}\cdot\mathbf{r}_{24}-e\mathbf{r}_{13}\cdot\mathbf{r}_{13,24}
-f\mathbf{r}_{24}\cdot\mathbf{r}_{13,24}},\nonumber
\end{eqnarray}
in which the coefficients $a$-$f$ depend on the variational
parameters $\nu$s and the elements of the transformation matrix from
\{$\mathbf{r}_{12}$, $\mathbf{r}_{34}$, $\mathbf{r}_{12,34}$\} to
\{$\mathbf{r}_{13}$, $\mathbf{r}_{24}$, $\mathbf{r}_{13,24}$\}. With
the cumbersome angular momentum algebra~\cite{vijande2009}, one can
show that the wave function
$\phi'(\mathbf{r}_{13},\mathbf{r}_{24},\mathbf{r}_{13,24})$ includes
all possible relative orbital angular momenta $l_{13}$, $l_{24}$ and
$l_{13,24}$ associated with the relative motions $\mathbf{r}_{13}$,
$\mathbf{r}_{24}$, and $\mathbf{r}_{13,24}$, respectively, which
should satisfy $l_{13}\oplus l_{24}\oplus l_{13,24}=0$ or 2 and
$l_{13}+l_{24}+l_{13,24}=even$ due to the parity conservation.
The details of the wave function can be found in the supplemental
materials at the bottom of this paper. After accomplishing the
above procedures, the diquark $[Q_1Q_3]$ and the antidiquark
$[\bar{q}_2\bar{q}_4]$ are either symmetric or antisymmetric.
Then, one can apply the Pauli principle to two pairs of identical
quarks, and naturally arrive at the diquark-antidiquark structure.

In fact, the above procedures are just making a rough partial
wave analysis on the eigenvector of the ground molecular state
$[Q_1\bar{q}_2][Q_3\bar{q}_4]$ according to the Pauli principle,
which deserves further investigation. Higher partial wave components
are less important from the nucleon-nucleon scattering.

\begin{table}[t]
\caption{Components, ratio, and binding energy $E_b$.}
\begin{tabular}{cccccccccccccccccc}
\toprule[0.8pt]
\noalign{\smallskip}
State&~~~~~~&~~Component~~&~~~~~~$E_b$ (MeV)~~~~~~&Ratio (~\%) \\
\noalign{\smallskip}
\toprule[0.8pt]
\noalign{\smallskip}
$T^+_{cc}$&& $D^*D$/$D^*D^*$ &$-1.2\pm0.2$&$74.2:6.8$\\
\noalign{\smallskip}
&&$D_{\mathbf{8}}^*D_{\mathbf{8}}$/$D_{\mathbf{8}}^*D_{\mathbf{8}}^*$&&$0.2:18.8$\\
\noalign{\smallskip}
$T^{+'}_{cc}$&& $D^*D$/$D^*D^*$& $-0.9\pm0.2$&$98.8:1.2$\\
\noalign{\smallskip}
$T^-_{bb}$&& $\bar{B}^*\bar{B}$/$\bar{B}^*\bar{B}^*$ &$-48.9\pm0.2$&$49.0:40.0$\\
\noalign{\smallskip}
 && $\bar{B}_{\mathbf{8}}^*\bar{B}_{\mathbf{8}}$/
$\bar{B}_{\mathbf{8}}^*\bar{B}_{\mathbf{8}}^*$&&$7.0:4.0$\\
\noalign{\smallskip}
$T^{-'}_{bb}$&& $\bar{B}^*\bar{B}$/$\bar{B}^*\bar{B}^*$ &$-43.8\pm0.2$&$62.4:37.6$\\
\noalign{\smallskip}
\toprule[0.8pt]
\end{tabular}
\caption{Probability of various combinations of $[Q_1Q_3]$ and
$[\bar{q}_2\bar{q}_4]$ in the state $[Q_1Q_3][\bar{q}_2\bar{q}_4]$
and their energy in unit of MeV. Orbit stands for the angular
momenta $l_{13}$, $l_{24}$ and $l_{13,24}$.}
\begin{tabular}{cccccccccccccccccc}
\toprule[0.8pt]
\noalign{\smallskip}
Case&~Color~&Spin&Orbit&~~~$[c_1c_3][\bar{q}_2\bar{q}_4]$~~~&$[b_1b_3][\bar{q}_2\bar{q}_4]$\\
\noalign{\smallskip}
\toprule[0.8pt]
\noalign{\smallskip}
(1)&$\bar{\mathbf{3}}\otimes\mathbf{3}$&$\mathbf{1}\oplus\mathbf{0}$&000&3820, 98.7\%&10355, 99.8\% \\
\noalign{\smallskip}
(2)&$\mathbf{6}\otimes\bar{\mathbf{6}}$&$\mathbf{0}\oplus\mathbf{1}$&000&4112, 1.3\%&10679, 0.2\%\\
\noalign{\smallskip}
&(1)+(2)&&&$3817\pm12$&$10355\pm13$\\
\noalign{\smallskip}
(3)&$\bar{\mathbf{3}}\otimes\mathbf{3}$&$\mathbf{0}\oplus\mathbf{1}$&110&4439, 1.5\%&10973, 0.3\%\\
\noalign{\smallskip}
(4)&$\mathbf{6}\otimes\bar{\mathbf{6}}$&$\mathbf{1}\oplus\mathbf{0}$&110&4285, 98.5\%&10841, 99.7\%\\
\noalign{\smallskip}
&(3)+(4)&&&$4282\pm8$&$10840\pm8$\\
\noalign{\smallskip}
(5)&$\bar{\mathbf{3}}\otimes\mathbf{3}$&$\mathbf{1}\oplus\mathbf{1}$&011&$4440\pm8$&$10987\pm8$\\
\noalign{\smallskip}
(6)&$\mathbf{6}\otimes\bar{\mathbf{6}}$&$\mathbf{1}\oplus\mathbf{1}$&101&$4311\pm9$&$10873\pm8$\\
\noalign{\smallskip}
\toprule[0.8pt]
\end{tabular}
\caption{Probability of various combinations of $[Q_1Q_3]$ and $[\bar{q}_2\bar{q}_4]$ in the state
$[Q_1\bar{q}_2][Q_3\bar{q}_4]$ with $01^+$.}
\begin{tabular}{cccccccccccccccccc}
\toprule[0.8pt]
\noalign{\smallskip}
Case&~~Color~~&Spin&Orbit&~~~$T^+_{cc}$~~~&$T^{+'}_{cc}$&~~~$T^-_{bb}$~~~&$T^{-'}_{bb}$\\
\noalign{\smallskip}
\toprule[0.8pt]
\noalign{\smallskip}
(a)&$\bar{\mathbf{3}}\otimes\mathbf{3}$&$\mathbf{1}\oplus\mathbf{0}$&$e$$e$$e$&13.6\%&8.4\%&13.4\%&11.0\%\\
\noalign{\smallskip}
(b)&$\mathbf{6}\otimes\bar{\mathbf{6}}$&$\mathbf{0}\oplus\mathbf{1}$&$e$$e$$e$&17.8\%&16.9\%&22.8\%&22.0\%\\
\noalign{\smallskip}
(c)&$\bar{\mathbf{3}}\otimes\mathbf{3}$&$\mathbf{0}\oplus\mathbf{1}$&$o$$o$$e$&13.6\%&8.4\%&13.4\%&11.0\%\\
\noalign{\smallskip}
(d)&$\mathbf{6}\otimes\bar{\mathbf{6}}$&$\mathbf{1}\oplus\mathbf{0}$&$o$$o$$e$&17.8\%&16.9\%&22.8\%&22.0\%\\
\noalign{\smallskip}
(e)&$\bar{\mathbf{3}}\otimes\mathbf{3}$&$\mathbf{1}\oplus\mathbf{1}$&$e$$o$$o$&12.4\%&16.5\%&10.5\%&11.3\%\\
\noalign{\smallskip}
(f)&$\mathbf{6}\otimes\bar{\mathbf{6}}$&$\mathbf{1}\oplus\mathbf{1}$&$o$$e$$o$&24.8\%&33.0\%&17.5\%&22.7\%\\
\noalign{\smallskip}
\toprule[0.8pt]
\end{tabular}
\end{table}

In general, a multiquark state should be a mixture of all possible
color components, including the color singlets and hidden color
configurations. Taking the $T^+_{cc}$ state with $01^+$ as an
example, it contains two color singlets $D^*D$ and $D^*D^*$ and
two hidden color octets $D_\mathbf{8}^*D_\mathbf{8}$ and
$D_\mathbf{8}^*D^*_\mathbf{8}$. However, the deuteron-like
configuration of the $T^+_{cc}$ indicates that the hidden color
components do not play critical role because the color confinement
interaction is suppressed in the distance larger than 1 fm when we
adopt the screened confinement potential~\cite{deng2022}.

The states with various components and the percentage of each
component are listed in Table I. One can find that the colorless
components in the $T^+_{cc}$ state occupy 81\% while the percentage
of the hidden color components is 19\%. The distance between the two
subclusters is 2.62 fm, which clearly supports the deuteron-like
configuration. In addition, we also give numerical results when the
state contains the color singlet components $DD^*$ and $D^*D^*$
only. The difference between the two cases is very small, which also
holds for the state $T^-_{bb}$. These numerical results indeed
justify the consideration of all possible colorless components only
in the deuteron-like configuration~\cite{deng2022}. On the other
hand, the hidden color component is an inevitable new degree of
freedom of the multi-quark states which is absent in ordinary
hadrons. The physical effects of the hidden color components
especially in the deeply bound multiquark states are interesting and
deserve further investigation.

The ground state $[Q_1Q_3][\bar{q}_2\bar{q}_4]$ with $01^+$ only
contains two channels, see the cases (1) and (2) in Table II, which
are the generally adopted diquark-antidiquark configurations. After
coupling the two channels, the two states
$[c_1c_3][\bar{q}_2\bar{q}_4]$ and $[b_1b_3][\bar{q}_2\bar{q}_4]$
with $01^+$ can generate deep compact bound states with the binding
energy of about $-60$ MeV and $-205$ MeV, respectively. The binding
energy mainly come from the contributions of the strong Coulomb
interaction, color-magnetic interaction and meson exchange
interaction. The optima spin-isospin-color-orbit combination (i.e.,
the case (1)) is dominant with a probability almost reaching 100\%,
so that the coupled channel effect is insignificant in the cases (1)
and (2). In addition, we present four lowest excited states with
$01^+$, which are much higher than the ground states. The cases (3)
and (4) have the same orbital excitation mode, where the coupled
channel effect is also very weak and the color configuration
$\mathbf{6}\otimes\bar{\mathbf{6}}$ is dominant. The cases (5) and
(6) can not couple with each other due to their different excitation
modes and the absence of noncentral forces in the
model~\cite{deng2022}. Furthermore, we find that the coupled channel
effect is always very weak between the same orbital excitation modes
after checking some higher excited states.

In order to illustrate the underlying mechanism of these two
different physical pictures clearly, we decompose the ground state
$[Q_1\bar{q}_2][Q_3\bar{q}_4]$ with $01^+$ as the superposition of a
large number of the $[Q_1Q_3][\bar{q}_2\bar{q}_4]$ states with
various angular excitations of $l_{13}$, $l_{24}$ and $l_{13,24}$.
According to the spin-color-isospin symmetry, six types of the
$[Q_1Q_3]$ and $[\bar{q}_2\bar{q}_4]$ combinations can satisfy the
Pauli principle and the parity conservation, see Table III. $e$ and
$o$ represent even and odd, respectively. For example, the $eee$
represents $000$, 020, 220, 222, 224, etc., which are orthogonal
bases and do not mix with each other because of the lack of
noncentral force in the model~\cite{deng2022}. With the probability
of each component in the states in Table I and their color-spin wave
functions, we can approximately exhibit the percentage of each type
of the $[Q_1Q_3]$ and $[\bar{q}_2\bar{q}_4]$ combination in the
states if we ignore the very weak coupled channel effect among the
modes with the same orbital excitations, which are presented in
Table III.

In principle, we can also expand the $[Q_1Q_3][\bar{q}_2\bar{q}_4]$
state in terms of the molecular $[Q_1\bar{q}_2][Q_3\bar{q}_4]$ bases.
On the contrary, the Pauli principle does not act on the molecules
$[Q_1\bar{q}_2]$ and $[Q_3\bar{q}_4]$ so that their orbitally
excited modes $l_{12}$ and $l_{34}$ can not be determined precisely
as the identical $[Q_1Q_3]$ and $[\bar{q}_2\bar{q}_4]$ in Table III.
There are many different combinations of the orbitally excited mode
(parity) for each isospin-spin-color combination if one attempts
to expand the $[Q_1Q_3][\bar{q}_2\bar{q}_4]$ state in terms of
the molecular $[Q_1\bar{q}_2][Q_3\bar{q}_4]$ bases. In another
word, one can make such expansions only according to isospin-spin-color
combinations, which leads to the fact that it is difficult
to extract some accurate and transparent information on the
molecular $[Q_1\bar{q}_2][Q_3\bar{q}_4]$ configuration from the
diquark state $[Q_1Q_3][\bar{q}_2\bar{q}_4]$.

In addition, we calculate the errors of the percentage, mass and
binding energy of the double heavy tetraquark states in the two
configurations, which are introduced by the uncertainty of the
adjustable model parameters in the Minuit program. The errors of
the percentage of each configuration in three Tables are less than
0.1\%. The mass errors in Table II are around 10 MeV. Those of the
excited states (i.e., the cases (3)-(6)) are 8-9 MeV while those of
the ground states (i.e., the cases (1) and (2)) are 12-13 MeV. The
definition of binding energy, $E_b=E_4-E_{m_1m_2}$, can greatly reduce
the influence of the uncertainty of the energy of the double heavy
tetraquark state $E_4$ and its corresponding twomeson threshold
$E_{m_1m_2}$ on the binding energy. Therefore, the $E_b$ errors are
very small, about 0.2 MeV, in Table I.

The optima isospin-color-spin-orbit combination only exists in the
case (a) in Table III. The percentage of the optima combination in the
$T^+_{cc}$ ($T^-_{bb}$) is in fact less than 13.6\% (13.4\%) because
the case (a) contains the contributions from the other orbitally excited
components. Compared with the $T^{+'}_{cc}$ ($T^{-'}_{bb}$), the $T^{+}_{cc}$
($T^{-}_{bb}$) has a larger percentage of the favorable optima combination
and a larger binding energy. In strong contrast, the probability of the
optima combination is almost 100\% in the state
$[c_1c_3][\bar{q}_2\bar{q}_4]$ ($[b_1b_3][\bar{q}_2\bar{q}_4]$), which
provides a strong attraction to generate a deep bound state. The optima
combination should therefore be responsible for the formation of the
shallow bound states from the prospective of diquarks.

The corresponding SU(2) groups of the isospin, $V$-spin and $U$-spin
are  three subgroups of the flavor SU(3) group. Therefore, the isospin,
$V$-spin, and $U$-spin antisymmetrical states, such as $T^-_{bb}$ with
$01^+$ and $T^-_{bbs}$ with $\frac{1}{2}1^+$, should share the same
symmetry in their wave functions so that their behaviors should be
analogous. For the state $T^-_{bbs}$ with $\frac{1}{2}1^+$, the
molecule configuration can form a stable state with a binding
energy of about $-10$ MeV, which implies the existence of the
deep bound diquark configuration $[b_1b_3][\bar{u}_2\bar{s}_4]$.
Its binding energy is about $-50$ MeV in the present calculation.
However, the state $T^+_{ccs}$ with $\frac{1}{2}1^+$ can not
produce a stable state in either molecule or diquark configuration.
The optima isospin ($V$-ispin)-color-spin-orbit combination is just
a necessary condition to produce a stable double heavy tetraquark
state. As a result, it is easy to understand why there do not exist
the stable doubly heavy tetraquark states which contain the isospin,
$U$-spin or $V$-spin symmetric light antidiquarks in various
theoretical frameworks.

The compact structure of the ground state
$[c_1c_3][\bar{q}_2\bar{q}_4]$ arises from the strong attractive
interaction within the $[c_1c_3]$ and $[\bar{q}_2\bar{q}_4]$
diquarks and the strong color force between the $[c_1c_3]$ and
$[\bar{q}_2\bar{q}_4]$. In contrast, in the state
$[c_1\bar{q}_2][c_3\bar{q}_4]$, the majority of the bad diquarks
with various orbitally excited modes are spatially extended while
the $c_1$ and $\bar{q}_2$ ($c_3$ and $\bar{q}_4$) are tightly bound
into a meson by the strong color force, which leads to the loosely
bound molecular state.

In principle, if all possible orbital excitations are considered
properly, either the diquark or molecule bases are complete and
orthogonal so that one can apply either set of these bases to make
the model calculations. One can decompose the complete bases into
two subsets: one is for the ground states and the other for the
excited states. Note that the corresponding subsets in the diquark
and molecule bases are not equivalent. The wave function of the
ground state does not mix with those of the orbitally excited states
if the model Hamiltonian does not contain the noncentral forces such
as the spin-orbital and $S$-$D$ wave mixing interactions as in
Ref.~\cite{deng2022}. The ground state bases with the diquark or
molecule configurations alone are enough to describe its
corresponding physical state in the model without the noncentral
interactions. On the other hand, the diquark and molecule
configurations have different orbitally excited modes. The
construction of the trial wave function in the model space depends
on the orbitally excited modes in the realistic calculations, which
may result in the difference between two configurations.

The existence of the deuteron-like $T^+_{cc}$ state implies the
advent of other compact double heavy tetraquark states. The doubly
charmed baryon $\Xi^{++}_{cc}$ indicates the possible existence of
the similar doubly charmed hadron with the light quark replaced by a
strongly correlated light anti-diquark. These interesting states may
also be searched for in the relativistic heavy-ion collisions at
ultrarelativistic energies~\cite{fries2008,zhang2021}.

\end{spacing}

\vspace{0.6cm}

\textbf{Conflict of interest}

\vspace{0.2cm}

The authors declare that they have no conflict of interest.

\acknowledgments
{One of the authors C. Deng thanks Prof. J.L. Ping for helpful
discussions. This research is partly supported by the National
Science Foundation of China under Contracts No. 11975033 and
No. 12070131001, Chongqing Natural Science Foundation under
Project No. cstc2019jcyj-msxmX0409 and Fundamental Research
Funds for the Central Universities under Contracts No. SWU118111.}

\begin{widetext}
\section*{Supplemental materials}
Completely equivalent color configurations
\{$\mathbf{1}_{12}\otimes\mathbf{1}_{34}$, $\mathbf{8}_{12}\otimes\mathbf{8}_{34}$\}
and \{$\bar{\mathbf{3}}_{13}\otimes\mathbf{3}_{24}$,
$\mathbf{6}_{13}\otimes\bar{\mathbf{6}}_{24}$\},
\begin{eqnarray}
\left (
\begin{array}{cccccc}
 \mathbf{1}_{12}\otimes\mathbf{1}_{34}   \\
 \noalign{\smallskip}
 \mathbf{8}_{12}\otimes\mathbf{8}_{34}    \\
\end{array}
\right )&=&\left (
\begin{array}{cccccc}
\frac{1}{\sqrt{3}}   & \frac{\sqrt{2}}{\sqrt{3}}             \\
\noalign{\smallskip}
-\frac{\sqrt{2}}{\sqrt{3}}  & \frac{1}{\sqrt{3}}              \\
\end{array}
\right )\left (
\begin{array}{cccccc}
\bar{\mathbf{3}}_{13}\otimes\mathbf{3}_{24}   \\
\noalign{\smallskip}
\mathbf{6}_{13}\otimes\bar{\mathbf{6}}_{24}
\end{array}
\right ).\nonumber
\end{eqnarray}
Completely equivalent spin configurations
\{$\mathbf{1}_{12}\oplus\mathbf{1}_{34}$,
$\mathbf{1}_{12}\oplus\mathbf{0}_{34}$,
$\mathbf{0}_{12}\oplus\mathbf{1}_{34}$\} and
\{$\mathbf{1}_{13}\oplus\mathbf{1}_{24}$,
$\mathbf{1}_{13}\oplus\mathbf{0}_{24}$,
$\mathbf{0}_{13}\oplus\mathbf{1}_{24}$\},
\begin{eqnarray}
\left (
\begin{array}{cccccc}
\mathbf{1}_{12}\oplus\mathbf{1}_{34}   \\
 \noalign{\smallskip}
\mathbf{1}_{12}\oplus\mathbf{0}_{34}    \\
\noalign{\smallskip}
\mathbf{0}_{12}\oplus\mathbf{1}_{34}    \\
\end{array}
\right )&=&\left (
\begin{array}{cccccc}
         0            & \frac{1}{\sqrt{2}} & \frac{1}{\sqrt{2}}            \\
\noalign{\smallskip}
-\frac{1}{\sqrt{2}}   &     \frac{1}{2}    & -\frac{1}{2}         \\
\noalign{\smallskip}
-\frac{1}{\sqrt{2}}   &     -\frac{1}{2}   &  \frac{1}{2}         \\
\end{array}
\right )\left (
\begin{array}{cccccc}
\mathbf{1}_{13}\oplus\mathbf{1}_{24}   \\
 \noalign{\smallskip}
\mathbf{1}_{13}\oplus\mathbf{0}_{24}    \\
\noalign{\smallskip}
\mathbf{0}_{13}\oplus\mathbf{1}_{24}    \\
\end{array}
\right ).\nonumber
\end{eqnarray}
A unitary transformation matrix
\begin{eqnarray}
\left (
\begin{array}{cccccc}
 \mathbf{r}_{12}   \\
 \noalign{\smallskip}
 \mathbf{r}_{34}    \\
 \noalign{\smallskip}
 \mathbf{r}_{12,34}      \\
\end{array}
\right )&=&\left (
\begin{array}{cccccc}
\frac{m_3}{m_1+m_3}   & \frac{-m_4}{m_2+m_4}    &        1         \\
\noalign{\smallskip}
\frac{-m_1}{m_1+m_3}  & \frac{m_2}{m_2+m_4}     &        1         \\
\noalign{\smallskip}
\frac{m_1m_3(m_1+m_2+m_3+m_4)}{(m_1+m_2)(m_1+m_3)(m_3+m_4)} & \frac{m_2m_4(m_1+m_2+m_3+m_4)}
{(m_1+m_2)(m_2+m_4)(m_3+m_4)} &\frac{m_1m_4-m_2m_3}{(m_1+m_2)(m_3+m_4)}       \\
\end{array}
\right )\left (
\begin{array}{cccccc}
\mathbf{r}_{13}   \\
\noalign{\smallskip}
\mathbf{r}_{24}    \\
\noalign{\smallskip}
\mathbf{r}_{13,24}
\end{array}
\right ).\nonumber
\end{eqnarray}
Gaussian wave function
\begin{eqnarray}
\psi^G_{lm}(\mathbf{x})=\sum_{n=1}^{n_{max}}c_{n}N_{nl_x}x^{l_x}e^{-\nu_{n}x^2}
Y_{l_xm}(\hat{\mathbf{x}}),\nonumber\\
\phi(\mathbf{r}_{12},\mathbf{r}_{34},\mathbf{r}_{12,34})=
\psi^G_{00}(\mathbf{r}_{12})\psi^G_{00}(\mathbf{r}_{34})
\psi^G_{00}(\mathbf{r}_{12,34}).\nonumber
\end{eqnarray}
Wave function expansion
\begin{eqnarray}
&e^{-d\mathbf{r}_{13}\cdot\mathbf{r}_{24}-e\mathbf{r}_{13}\cdot\mathbf{r}_{13,24}
-f\mathbf{r}_{24}\cdot\mathbf{r}_{13,24}}=\frac{1}{4\sqrt{\pi}}\sum_{l_{13}=0}^{\infty}
\sum_{l_{24}=0}^{\infty}\sum_{l_{13,24}=0}^{\infty}
\left[[Y_{l_{13}}(\hat{\mathbf{r}}_{13})Y_{l_{24}}(\hat{\mathbf{r}}_{24})]_{l_{13,24}}
Y_{l_{13,24}}(\hat{\mathbf{r}}_{13,24})\right]_0\nonumber\\
&\times\sum_{l_1,l_2,l_3}\left(2l_1+1\right)\left(2l_2+1\right)\left(2l_3+1\right)\langle l_10l_20|l_{13}\rangle
\langle l_10l_30|l_{24}\rangle\langle l_20l_30|l_{13,24}\rangle\left \{
\begin{array}{cccccc}
l_{13} &  l_{24} &  l_{13,24} \\
\noalign{\smallskip}
l_3    &  l_2    &  l_1       \\
\end{array}
\right\}\nonumber\\
&\times\left(\sqrt{\frac{\pi}{2dr_{13}r_{24}}}I_{l_1+\frac{1}{2}}(dr_{13}r_{24})\right)
\left(\sqrt{\frac{\pi}{2er_{13}r_{13,24}}}I_{l_2+\frac{1}{2}}(er_{13}r_{13,24})\right)
\left(\sqrt{\frac{\pi}{2fr_{24}r_{13,24}}}I_{l_3+\frac{1}{2}}(fr_{24}r_{13,24})\right),\nonumber
\end{eqnarray}
where $I_a(x)$ are the modified Bessel functions.
\end{widetext}
\end{document}